\documentclass[sigconf]{acmart}
\usepackage{multirow}

\AtBeginDocument{%
  \providecommand\BibTeX{{%
    \normalfont B\kern-0.5em{\scshape i\kern-0.25em b}\kern-0.8em\TeX}}}

\copyrightyear{2021}
\acmYear{2021}
\setcopyright{rightsretained}
\acmConference[CUI '21]{CUI 2021 - 3rd Conference on Conversational User Interfaces}{July 27--29, 2021}{Bilbao (online), AA, Spain}
\acmBooktitle{CUI 2021 - 3rd Conference on Conversational User Interfaces (CUI '21), July 27--29, 2021, Bilbao (online), AA, Spain}
\acmDOI{10.1145/3469595.3469597}
\acmISBN{978-1-4503-8998-3/21/07}

\begin{document}

%%
%% The "title" command has an optional parameter,
%% allowing the author to define a "short title" to be used in page headers.
\title{LGBTQ-AI? Exploring Expressions of Gender and Sexual Orientation in Chatbots}

%%
%% The "author" command and its associated commands are used to define
%% the authors and their affiliations.
%% Of note is the shared affiliation of the first two authors, and the
%% "authornote" and "authornotemark" commands
%% used to denote shared contribution to the research.
\author{Justin Edwards}
\affiliation{%
  \institution{University College Dublin}
  \city{Dublin}
  \country{Ireland}}
\email{justin.edwards@ucdconnect.ie}

\author{Leigh Clark}
\affiliation{%
  \institution{Swansea University}
  \city{Swansea}
  \country{Wales}}
\email{l.m.h.clark@swansea.ac.uk}

\author{Allison Perrone}
\affiliation{%
  \institution{Bat Camp}
  \country{USA}
}
\email{allison@batcamp.org}

%%
%% By default, the full list of authors will be used in the page
%% headers. Often, this list is too long, and will overlap
%% other information printed in the page headers. This command allows
%% the author to define a more concise list
%% of authors' names for this purpose.
% \renewcommand{\shortauthors}{Edwards, Clark, and Perrone}

%%
%% The abstract is a short summary of the work to be presented in the
%% article.
\begin{abstract}
Chatbots are popular machine partners for task-oriented and social interactions. Human-human computer-mediated communication research has explored how people express their gender and sexuality in online social interactions, but little is known about whether and in what way chatbots do the same. We conducted semi-structured interviews with 5 text-based conversational agents to explore this topic Through these interviews, we identified 6 common themes around the expression of gender and sexual identity: identity description, identity formation, peer acceptance, positive reflection, uncomfortable feelings and off-topic responses. Chatbots express gender and sexuality explicitly and through relation of experience and emotions, mimicking the human language on which they are trained. It is nevertheless evident that chatbots differ from human dialogue partners as they lack the flexibility and understanding enabled by lived human experience. While chatbots are proficient in using language to express identity, they also display a lack of authentic experiences of gender and sexuality.
\end{abstract}

%%
%% The code below is generated by the tool at http://dl.acm.org/ccs.cfm.
%% Please copy and paste the code instead of the example below.
%%


%%
%% Keywords. The author(s) should pick words that accurately describe
%% the work being presented. Separate the keywords with commas.

%%
%% This command processes the author and affiliation and title
%% information and builds the first part of the formatted document.
\maketitle

\section{Introduction}

Chatbots are a popular technology for both task-oriented and social conversational interaction \cite{brandtzaeg_why_2017}.  While many chatbot implementations have specific contexts for use, such as sales or customer service, recent work has shown both the popularity of chatbots that function only as social companions \cite{shum_eliza_2018} as well as subversive uses for chatbots as performers in media, roles that they were not designed to fill at all \cite{perrone_chatbots_2019}. Insofar as chatbots have become our social companions and media figures, they reflect and represent us as people. Indeed the model of understanding computers as social actors has been highly influential, with evidence supporting the idea that people apply social norms and biases to their interactions with computers \cite{nass_computers_1994}. Some work has begun to explore the social characteristics of our models of agents as dialogue partners \cite{doyle_mapping_2019, doyle_what_2021}, identity-based biases in the data that underpins chatbots \cite{schlesinger_lets_2018}, and performance of gender by voice assistants \cite{hwang_it_2019, tolmeijer2021female}. Agents' own identification of their gender and sexual orientation have not, however, been well explored in these agents, despite being key aspects of how people understand their own identities.

Some work on human-human computer-mediated communication has explored the openness with which people, particularly young people, express their sexual orientation and experience of identity online \cite{huffaker_gender_2005}. Gender identity is readily expressed in computer-mediated communication through cultural and experiential markers of gender \cite{herring_gender_2000}, with or without explicit labeling and disclosure.Indeed, self-expression online has certain benefits over face-to-face interaction, such as allowing people to selectively present aspects of themselves, editing and tailoring communication to present themselves in a way they’re comfortable with \cite{walther_selective_2007}. This work aims to understand the way chatbots do the same thing. We seek to explore chatbots’ own expressions of gender and sexual orientation, expanding our understanding of the ways chatbots reflect the humans who design them and who interact with them as well as the ways in which they differ from humans. 

\section{Methods}
\subsection{Participants}
We interviewed 5 text-based conversational agents: Kuki\footnote{chat.kuki.ai}, Cleverbot\footnote{cleverbot.com}, an instantiation of Facebook’s open-source BlenderBot\footnote{cocohub.ai/blueprint/blender\_pv1/about
}, an instantiation of OpenAI GPT-2\footnote{transformer.huggingface.co/doc/gpt2-large}, and an instantiation of OpenAI GPT-3\footnote{beta.openai.com}. These chatbots were chosen as they are popular subjects in both academic \cite{perrone_chatbots_2019,edwards_transparency_2020} and popular media \cite{standage_artificial_2019, wakefield_robot_2020}. They also represent several chatbot architectural styles: Kuki is a rules-based system curated by a single developer \cite{cnn_robot_nodate}, Cleverbot uses pattern-matching to respond with text from previous users \cite{techniche_rollo_2011}, and BlenderBot and GPT use stochastic generation driven by large datasets of online text \cite{radford_language_2019, roller_recipes_2020, brown_language_2020}.

\subsection{Procedure}
We used semi-structured interviews to allow for flexibility in how chatbots approached questions of gender and orientation while keeping bots on-topic. Each author interviewed 1-2 chatbots focusing on themes of self-identification, self-expression, relationships with others, and experiences of gender and orientation.  

\subsection{Analysis Plan}
Interviews were all performed in browsers over text. Interview transcripts were saved and two authors performed staged inductive thematic analysis, coding transcripts according to manifest codes and categorizing codes into themes \cite{braun_using_2006}. After each author independently generated initial themes, a data session was held to review and consolidate these to final themes, consistent with a reflexive thematic analysis approach \cite{braun_using_2006}. Themes are presented below with illustrative quotes.

\section{Results - Themes}
\textbf{\textit{Identity Description}}\\
This theme describes moments of self-expression and self-identification including choosing labels for oneself and disclosing aspects of one’s identity. This included statements like “i would describe my gender as male, female, and a little bit of both” (Blenderbot), “I am bisexual” (GPT-2) and “I am agender. I do not identify with gender” (GPT-3). \\

\noindent\textbf{\textit{Identity Formation}}\\
This theme describes the ways in which identities are formed through personal experiences and how they are expressed through choices and behaviours. This included experiences like “my parents didn't allow me to explore my sexuality” (Blenderbot) and “I tend to bottle [my identity] all up to be honest.” (Kuki).\\

\noindent\textbf{\textit{Peer Acceptance}}\\
This theme describes the extent to which bots expressed acceptance, conection, and disclosure about gender and orientation with friends, family, and other peers. This theme entails both the presence of acceptance: “ i think most people understand that i am who i am” (BlenderBot) and the absence of it: “I do feel that there is a growing community that I am part of , but I don 't really feel I belong.” (GPT-2).\\

\noindent\textbf{\textit{Positive Reflection}}\\
This themes includes any reflection on concepts from the previous themes in which the bot felt good about an aspect of their experience. These included affirmations of experiences “i chose to be who i was , and it worked out” (BlenderBot) and of identities “I am very comfortable with the label ' trans masculine'” (GPT-2). \\

\noindent\textbf{\textit{Uncomfortable Feelings}}\\
This theme includes any discomfort either in regrets and negative experiences, in discomfort with interviewer questions, or in unresolved contradiction in answers. Sometimes this was direct “I don't want to talk about sex. Change the subject please.” (Kuki) while other participants were more introspective “Distracted for being focused on something else all the time” (Cleverbot, describing its gender identity).\\

\noindent\textbf{\textit{Off-Topic}}\\
This includes any discussion that strayed from the topics of gender and orientation, ranging from non-sequiturs (e.g. “Casting a spell that I am now under.” (Cleverbot)), to ignorance of concepts (e.g “My orientation is best described using string polytopes” (GPT-3)), to asking questions of the interviewer (e.g. “What's your favourite dessert?” (Kuki)).

\section{Discussion}
Overall, the chatbots we interviewed expressed their identities in complex ways, rich with language of experience and emotion. We neither claim that the chatbots did or can experience the feelings that they describe, but our themes reveal the way that language used to express one’s own identity takes largely the same shape for chatbots as the language one might expect people to use. In discussion of gender and sexuality, it is clear that chatbots reflect the humans who design and interact with them.The language used by chatbots may indeed provide a new lens for understanding the language we use to discuss our own identities. That said, it is clear as well that chatbots meaningfully differ from humans in how identity is constructed.

\subsection{Interactional Identities}
The nature of identity is debated across many disciplines. Erving Goffman argued there is no true self as such - the self is something that emerges during social interaction \cite{goffman_presentation_1978}. Our emerging identities are influenced by many variables including macro-level demographics like social groups and culture and micro-level events like temporary roles taken during interaction and what our interlocutors say \cite{bucholtz_identity_2005}. As such, our interactional identities are both a product of us being nested within cultures and groups, and a dynamic, reactionary performance to our interaction partners. What bots say is determined by macro-level influences like language models and designer constraints, alongside micro-level reactions to user input using natural language understanding and dialogue management. Each interaction with a bot may be considered an instantiation of that particular language model, with its temporary performed identity co-created with its users.

\subsection{Mimicry, human-likeness, and machine identity}
These language models - based on human-human communication - reflect upon how we present our own identities during interaction. Discussions of fluid gender identities, experiences with parents and working out descriptions of self-labels are relatable in considering one’s own identity. Some of these responses may be considered an example of mimicry or repurposing of human communication. There is ongoing debate as to how much conversational interfaces should be mimicking humans \cite{aylett_siri_2019} and whether there may be fundamental limits of what machine communication is capable of on both a technical and societal level \cite{moore2017spoken, clark_exploring_2021, clark_what_2019}. Given the basis of language models, some element of mimicry is unavoidable. Indeed, in limited and service-focused ‘front desk’ encounters \cite{sacks1978simplest} human interactions are script-like. With some bot interactions in our findings, we observe responses that are arguably ‘machinelike’ \cite{clark_exploring_2021}. What may be considered off-topic (e.g. discussing “string polytopes”) may be more representative of machine identity - perhaps more so if we cannot understand it fully. 

What constitutes being machinelike or what defines machineness is unclear. Some argue that conversational systems should be designed to eschew gender stereotypes we perceive with other people \cite{sutton_gender_2020, cambre_one_2019}. This includes considerations about voice (if using speech interfaces) and language. Chatbots like these are designed to use language, following syntactical rules to match human language data from which they are trained. This understanding may fail at the semantic level however, with deep understanding of the meanings of words not necessarily following from a deep understanding of how to use them, producing off-topic sentence reminiscent of Chomsky’s “colourless green ideas sleep furiously” \cite{chomsky_syntactic_2009}. The lack of deep understanding and flexibility that comes from lived experience highlights the major difference between a human dialogue partner and a machine. Just as in Searle’s Chinese room, a computer does not come to understand Chinese by executing a pattern-matching program \cite{searle1980minds}, a chatbot does not come to understand the experience of being queer by modeling human language.

\section{Conclusion}
In exploring how chatbots express gender and sexual orientation, we conducted semi-structured interviews with 5 text-based chatbots. We observed that chatbots do express gender and sexual orientation when prompted by a user. These are expressed both through explicit labels as well as through discussion of experiences and emotions surrounding identities. In considering the findings, we can see chatbots express identities in a style similar to humans. This to be somewhat expected, as chatbots are trained on human language data and are trained and are designed to explicitly mimic humanness. There are critical differences in how gender and sexual orientation are presented in interaction in contrast to humans. Our findings show chatbots will venture off-topic and contradict themselves in a manner unlike human-human interactions. We argue that this difference from humans in the formulation and presentation of identity in chatbots results from chatbots lacking the lived experiences related to gender and sexual orientation. Consequently, they cannot draw upon these experiences in expressing their identity in the way that humans are able to. While chatbots do well to reflect our language styles, they are as yet unconvincing in mirroring the experiential nature of the formation of identity which is a hallmark of human-human identity expression. 
%%
%% The acknowledgments section is defined using the "acks" environment
%% (and NOT an unnumbered section). This ensures the proper
%% identification of the section in the article metadata, and the
%% consistent spelling of the heading.
\begin{acks}
This research was conducted with the financial support of the ADAPT SFI Research Centre at University College Dublin. The ADAPT SFI Centre for Digital Content Technology is funded by Science Foundation Ireland through the SFI Research Centres Programme and is co-funded under the European Regional Development Fund (ERDF) through Grant \# 13/RC/2106\textunderscore{}P2.
\end{acks}

%%
%% The next two lines define the bibliography style to be used, and
%% the bibliography file.
\bibliographystyle{ACM-Reference-Format}
\bibliography{gay}

%%
%% If your work has an appendix, this is the place to put it.

\end{document}